\documentclass[twocolumn,showpacs,aps,prd,superscriptaddress]{revtex4}

\usepackage{graphicx}
\usepackage{units}

\input{pubboard/babarsym}
\newcommand{\Bztokstargamma}{\ensuremath{\Bz\to\Kstarz\gamma}}

\newcommand{\Bztokspizgamma}{\ensuremath{\Bz\to\KS\piz\gamma}}
\newcommand{\Kstartokspiz}{\ensuremath{\Kstarz\to\KS\piz}}
\newcommand{\skspizgamma}{\ensuremath{S_{\KS\piz\gamma}}}
\newcommand{\ckspizgamma}{\ensuremath{C_{\KS\piz\gamma}}}
\newcommand{\skstargamma}{\ensuremath{S_{\Kstarz\gamma}}}
\newcommand{\ckstargamma}{\ensuremath{C_{\Kstarz\gamma}}}
\newcommand{\sbb}{\ensuremath{S_{\BB}}}
\newcommand{\cbb}{\ensuremath{C_{\BB}}}

\newcommand{\mkspiz}{\ensuremath{m_{\KS\piz}}}
\newcommand{\Bflav} {\ensuremath{B_{\rm flav}}}

\newcommand{\Brec}{\ensuremath{B_{\CP}}}
\newcommand{\Btag}{\ensuremath{B_\mathrm{tag}}}
\newcommand{\cosThetaB}{\ensuremath{\cos\!\theta_B^*}}
\newcommand{\cosThetaK}{\ensuremath{\cos\!\theta_{K^*}}}
\newcommand{\dte}{\ensuremath{\sigma(\deltat)}}

\newcommand{\tauB}{\ensuremath{\tau_{B}}}
\newcommand{\belle}{BELLE}

\newcommand{\BABARPubYear}    {05}
\newcommand{\BABARPubNumber}  {30}

\newcommand{\SLACPubNumber} {11345}

\def\figurebox#1#2#3{%
    \def\arg{#3}%
    \ifx\arg\empty
    {\hfill\vbox{\hsize#2\hrule\hbox to #2{\vrule\hfill\vbox to #1{\hsize#2\vfill}\vrule}\hrule}\hfill}%
    \else
    {\hfill\epsfbox{#3}\hfill}%
    \fi}

\long\def\inst#1{\par\nobreak\kern 4pt\nobreak
    {\it #1}\par\vskip 10pt plus 3pt minus 3pt}

\begin{document}


\begin{flushleft}
\babar-PUB-\BABARPubYear/\BABARPubNumber\\
SLAC-PUB-\SLACPubNumber\\
\end{flushleft}

\title{ \Large \bf \boldmath Measurement of the time-dependent
  \CP-violating asymmetry in \Bztokspizgamma{} decays}

%
\author{B.~Aubert}
\author{R.~Barate}
\author{D.~Boutigny}
\author{F.~Couderc}
\author{Y.~Karyotakis}
\author{J.~P.~Lees}
\author{V.~Poireau}
\author{V.~Tisserand}
\author{A.~Zghiche}
\affiliation{Laboratoire de Physique des Particules, F-74941 Annecy-le-Vieux, France }
\author{E.~Grauges}
\affiliation{IFAE, Universitat Autonoma de Barcelona, E-08193 Bellaterra, Barcelona, Spain }
% ...
\collaboration{The \babar\ Collaboration}
\noaffiliation

\begin{abstract}
  We present a measurement of the time-dependent \CP-violating
  asymmetry in \Bztokstargamma{} decays with \Kstartokspiz{} based on
  232 million $\Y4S\to\BB$ decays collected with the \babar\ detector
  at the PEP-II asymmetric-energy \epem{} collider at SLAC. In a
  sample containing $157\pm 16$ signal decays, we measure
  $\skstargamma = -0.21 \pm 0.40 \pm 0.05$ and $\ckstargamma = -0.40
  \pm 0.23 \pm 0.03$, where the first error is statistical and the
  second systematic. We also explore \Bztokspizgamma{} decays with
  \unit[$1.1<\mkspiz<1.8$]{\gevcc} and find $59\pm13$ signal events
  with $\skspizgamma = 0.9 \pm 1.0 \pm 0.2$ and $\ckspizgamma = -1.0
  \pm 0.5 \pm 0.2$.
\end{abstract}


\maketitle

The decay transition $b\to s \gamma$ is sensitive to contributions
from physics beyond the Standard Model (SM)~\cite{NewPhysics}.  There
has been extensive experimental and theoretical investigation of the
inclusive decay rate $\BR(B\to X_s \gamma)$, which to date shows
no significant deviation from the SM~\cite{ComparisonBF}.  Various new
physics scenarios can accommodate large deviations from the SM in
other $b\to s \gamma$ decay properties as well, in particular in
\CP{}-violating (CPV) asymmetries and the polarization of the final
state photon~\cite{soni}. The photon polarization in $b\to s \gamma$
($\bar b\to \bar s\gamma$) is predominantly left handed (right handed)
in the SM. As a consequence, in the exclusive decay $B^0\to
(K^0_s\pi^0)\gamma$ interference of the amplitude for the direct decay
and the amplitude for the decay via $\Bz-\Bzb$ mixing is suppressed. Therefore,
time-dependent \CP{}-violating asymmetry is expected
to be small~\cite{soni},
$\skstargamma\approx-2\frac{m_s}{m_b}\stwob\approx-0.04$, where $m_s$
($m_b$) is the mass of the $s$ ($b$) quark,
$\beta\equiv\arg(-V_{cd}V^*_{cb}/V_{td}V^*_{tb})$ and $V$ is the quark
mixing matrix~\cite{CKM}.  Any significant deviation that goes beyond
possible hadronization corrections of order
$0.1$~\cite{Grinstein:2004uu} would indicate phenomena beyond the SM.

In this Communication we report new measurements of the time-dependent CPV
asymmetry in \Bztokspizgamma{}~\cite{ref:cc} based on
$232$~million $\Y4S\to\BB$ decays collected with the \babar{} detector
at the \pep2{} asymmetric-energy \epem{} collider at SLAC.
Measurements of the CPV asymmetry in \Bztokstargamma{}, the subset of
events with $0.8<\mkspiz<1.0$, have previously been reported by
\babar{} on \unit[110]{\invfb}~\cite{Aubert:2004pe} and \belle{} on
\unit[253]{\invfb}~\cite{Ushiroda:2005sb}. The \belle{} collaboration
has also reported a measurement of inclusive \Bztokspizgamma{} with
\unit[$0.6<\mkspiz<1.8$]{\gevcc}~\cite{Ushiroda:2005sb}. The latter
measurement is motivated by a recent theoretical result that indicates
that all contributions to the $\KS\piz\gamma$ final state have the
same \CP{} eigenvalue~\cite{Atwood:2004jj}, so that beyond-the-SM
effects can be discovered even if the \mkspiz{} resonance structure is
not resolved.  Since the correctness of such an averaging procedure is
still under discussion~\cite{Grinstein:2004uu}, we present our results
for events with an invariant mass of the $\KS\piz$ pair near and above
the $K^{*}(892)^0$ resonance separately. For simplicity we refer
these two contributions as ``$K^{*}$'' and ``non-$K^{*}$'
respectively.

The \babar{} detector is fully described in Ref.~\cite{ref:babar}. The
components that are most important for this analysis are a five-layer
double-sided silicon micro-strip detector (SVT), a 40-layer drift
chamber (DCH) and a CsI(Tl) electromagnetic calorimeter (EMC).  For
event simulation we use the Monte Carlo event generator
EVTGEN~\cite{Lange:2001uf} and GEANT4~\cite{Agostinelli:2002hh}.

At the \Y4S{} resonance time-dependent CPV asymmetries are extracted
from the distribution of the difference of the proper decay times
$\deltat\equiv t_{\CP}-t_\mathrm{tag}$, where $t_{\CP}$ refers to the
 decay time of the signal \B{} (\Brec{}) and $t_\mathrm{tag}$ to
that of the other \B{} (\Btag). The \deltat{} distribution for
$\Brec\to f$ follows
\begin{eqnarray}
  \label{eqn:td}
  \lefteqn{{\cal P}_{\pm}(\deltat) \; = \; \frac{e^{-|\deltat|/\tauB}}{4\tauB} \; \times }\; \\
   && \left[ \: 1 \; \pm \;
    \: S_f \sin{( \deltamd\deltat)} \; \mp \; C_f \cos{( \deltamd\deltat)} \: \right] \; , \nonumber
\end{eqnarray}
where the upper (lower) sign corresponds to \Btag{} decaying as \Bz{}
(\Bzb), \tauB{} is the \Bz{} lifetime and \deltamd{} is the mixing
frequency. The coefficients $C_f$ and $S_f$ can be expressed in terms
of the \Bz-\Bzb{} mixing amplitude and the decay amplitudes for
$\Bz\to f$ and $\Bzb\to f$~\cite{lambda}. Direct \CP{} violation in
the decay $\Bz\to\ f$ results in a non-zero value of $C_f$. For
\Bztokstargamma{} direct \CP{} violation is constrained by
measurements of the partial rate asymmetry in decays with
$\Kstarz\to\Kp\pim$, ${\cal A}^{\CP}_{\Kstarz\gamma} =
-C_{\Kstarz\gamma} = -0.010 \pm
0.028$~\cite{HFAGACP}, which is in good agreement with the SM
prediction~\cite{Kagan:1998bh}.

We search for \Bztokspizgamma{} decays in \BB{} candidate events,
which are selected based on charged particle multiplicity and event
topology~\cite{ref:Sin2betaPRD}. Candidates for $\KS\to\pip\pim$ are
formed from pairs of oppositely charged tracks with a vertex $\chi^2$
probability larger than $0.001$, a $\pip\pim$ invariant mass
\unit[$487 < m_{\pip\pim} < 507$]{\mevcc} and a reconstructed decay
length greater than five times its uncertainty.  Photon candidates are
reconstructed from clusters in the EMC that are isolated from any
charged tracks and have the expected lateral shower shape. We form
$\piz\to\gamma\gamma$ candidates with an invariant mass \unit[$115 <
m_{\gamma\gamma} < 155$]{\mevcc} and energy
\unit[$E_{\piz}>590$]{\mev} from pairs of candidate photons each of
which carries a minimum energy of \unit[$30$]{\mev}. For the photon
from the $B$ decay, the so-called primary photon, we require
an energy in the \epem{} frame of \unit[$1.5 < E^*_\gamma
<3.5$]{\gev}. We veto primary photons that form a
$\piz\to\gamma\gamma$ ($\eta\to\gamma\gamma$) candidate with invariant
mass \unit[$115 < m_{\gamma\gamma} < 155$]{\mevcc}
(\unit[$470<m_{\gamma\gamma}<620$]{\mevcc}) when combined with another
photon with energy \unit[$E_\gamma>50$]{\mev}
(\unit[$E_\gamma>250$]{\mev}).

To identify \Bz{} decays in $\KS\piz\gamma$ combinations we use the
energy-substituted mass $\mes=\sqrt{(s/2+{\bf p}_i\cdot{\bf
    p}_B)^2/E_i^2-p^2_B}$ and the energy difference
$\DeltaE=E^*_B-\sqrt{s}/2$.  Here $(E_i,{\bf p}_i)$ and $(E_B,{\bf
  p}_B)$ are the four-vectors of the initial \epem{} system and the
$B$ candidate, respectively, $\sqrt{s}$ is the center-of-mass energy,
and the asterisk denotes the \epem{} rest frame.  For signal decays,
the \mes{} distribution peaks near the $B$ mass with a resolution of
about \unit[$3.5$]{\mevcc} and \DeltaE{} peaks near 0\mev{} with a resolution
of about \unit[$50$]{\mev}. Both \mes{} and \DeltaE{} exhibit a low-side tail
from energy leakage in the EMC.  We require $5.2<\mes<5.3\gevcc$ and
$|\DeltaE|<250\mev$, which includes the signal region as well as a
large ``sideband'' region for background estimation. We also require
$|\cosThetaB|<0.9$, where $\theta_B^{*}$ is the angle of the \B{}
candidate with respect to the $e^-$ momentum in the \epem{} rest
frame. Finally, for the subset of events with
\unit[$\mkspiz<1.1$]{\gevcc}, we require $|\cosThetaK|<0.9$, where
$\theta_{K^*}$ is the angle between the \KS{} and the primary photon
in the $\KS\piz$ rest frame (the ``helicity'' angle).

Event topology is exploited to further suppress the background from
continuum $\epem\to\qqbar$ ($q=u,d,s,c$) events.  We calculate
the ratio $L_{2}/L_{0}$ of two moments defined as
$L_j\equiv\sum_i |{\bf p}^*_i| |\cos \theta^*_i|^j$, where ${\bf
  p}^*_i$ is the momentum of particle $i$ in the \epem{} rest frame,
$\theta^*_i$ is the angle between ${\bf p}^*_i$ and the thrust axis of
the \B{} candidate and the sum runs over all reconstructed particles
except for the \B{} candidate daughters.  We require $L_2/L_0<0.55$,
which suppresses the background by more than a factor $3$ at the cost
of approximately \unit[$10$]{\%} signal efficiency. After all
selections are applied the average candidate multiplicity in events
with at least one candidate is approximately $1.1$.  We select the
candidate with a reconstructed \piz{} mass closest to the expected
value and if ambiguity persists we select the candidate with \KS{}
mass closest to the expected value.

Selected events are divided in events with
\unit[$0.8<\mkspiz<1.0$]{\gevcc}, where signal decays are
predominantly \Bztokstargamma{}, and events with
\unit[$1.1<\mkspiz<1.8$]{\gevcc}, where the contribution from
$K^*(892)$ is small. In the data we find respectively $1469$ and
$2629$ candidate events in these categories. The selection efficiency
for \Bztokstargamma{}, evaluated with simulated events, is
approximately \unit[$16$]{\%}.  Using the current world average for
the branching fraction~\cite{HFAGBF} we expect $176\pm18$ signal
events.  Compared to our previous measurement~\cite{Aubert:2004pe} the
current event selection is more effective in suppressing background
from $B$ decays, leading to a reduced systematic uncertainty from an
eventual CPV asymmetry in the background without a significant loss in
statistical sensitivity.  The selection efficiency for
\Bztokspizgamma{} events with \unit[$1.1<\mkspiz<1.8$]{\gevcc} is
approximately \unit[$15$]{\%}, but depends on the helicity structure.
Besides the $K^*(892)$ the only observed $K\pi$ resonance in $B\to
K\pi\gamma$ decays is the $K^*_2(1430)$. Using the world average for
the $\Bz\to K^*_2(1430)^0\gamma$ branching
fraction~\cite{HFAGBFkstartwo} we expect $24\pm7$ events. However,
since upper bounds on other resonances are weak, the actual observed
signal yield may be appreciably higher.

For each $B$ candidate we examine the remaining tracks
in the event to determine the decay vertex position and the flavor 
of \Btag. Using a neural network based on kinematic and particle
identification information~\cite{Aubert:2004zt} each event is assigned
to one of seven mutually exclusive tagging categories, designed to
combine flavor tags with similar performance and \deltat{} resolution.
We parameterize the performance of this algorithm in a data sample
(\Bflav{}) of fully reconstructed $\Bz\to D^{(*)-}\pip/\rho^+/a_1^+$
decays. The average effective tagging efficiency obtained from this
sample is $Q = \sum_c \epsilon_S^c (1-2w^c)^2=0.305\pm 0.004$, where
$\epsilon_S^c$ and $w^c$ are the efficiencies and mistag
probabilities, respectively, for events tagged in category
$c=1,\ldots7$.

The proper-time difference is extracted from the separation of the
\Brec{} and \Btag{} decay vertices in a manner analogous to
Ref.~\cite{Aubert:2004xf}. The \Btag{} vertex is reconstructed
from the remaining charged particles in the
event~\cite{ref:Sin2betaPRD}. To reconstruct the \Brec{} vertex from
the single \KS{} trajectory we exploit the knowledge of the average
interaction point (IP), which is determined from the spatial
distribution of vertices in two-track events and is calculated
separately for each 10-minute period of data-taking. We compute
\deltat{} and its uncertainty from a geometric
fit~\cite{Hulsbergen:2005pu} to the $\Y4S\to\Bz\Bzb$ system that takes
this IP constraint into account. We further improve the \deltat{}
resolution by constraining the sum of the two $B$ decay times
($t_{\CP}+t_\mathrm{tag}$) to be equal to $2\:\tau_{\Bz}$ with an
uncertainty $\sqrt{2}\; \tau_{\Bz}$. We have verified in a Monte-Carlo
simulation that this procedure provides an unbiased estimate of
\deltat{}.

The per-event estimate of the uncertainty on \deltat{} reflects the
strong dependence of the \deltat{} resolution on the $\KS$ flight
direction and on the number of SVT layers traversed by the $\KS$ decay
daughters. In about \unit[$70$]{\%} of the events both pion tracks are
reconstructed from at least 4 SVT hits, leading to sufficient
resolution for the time-dependent measurement. The average \deltat{}
resolution in these events is about \unit[$1.1$]{ps}. For events that
fail this criterion or for which \unit[$\dte>2.5$]{ps} or
\unit[$|\deltat|>20$]{ps}, the \deltat{} information is not used.
However, these events still contribute to the measurement of
\ckstargamma{}, which can also be extracted from flavor-tagging
information alone.

Signal yields and CPV asymmetries are extracted using an unbinned
maximum-likelihood fit to \mes{}, \DeltaE{}, $L_2/L_0$, flavor-tag,
\deltat{} and \dte{}, as in Ref.~\cite{Aubert:2004pe}.  For
the analysis of the \Bztokstargamma{} sample \mkspiz{} is also used in
the fit. Because we expect a contribution from other \B{} decays
(``\BB{} background''), we allow the fit to extract the fraction of such
decays as well. We have verified using fits to simulated samples that
the correlation between the observables is sufficiently small that the
event likelihoods for signal ${\cal P}_S$, \BB{} background ${\cal
  P}_{\BB}$ and continuum background ${\cal P}_{\qqbar}$ can be
described by the product of one-dimensional probability density
functions (PDF).  The PDFs for signal events and \BB{} background
events are parameterized using either the \Bflav{} sample (for the
flavor-tag efficiency, mistag probabilities and \deltat{}-resolution
function) or simulated events.  For the continuum background, we
select the functional form of the PDFs in background-enhanced samples.
We exploit the large fraction of background events in the final sample
to extract the background parameters along with the physics
measurements in the fit.  The asymmetry in the rate of $\Bz$ versus
$\Bzb$ tags in background events is also extracted from in the fit.

The PDF for the \deltat{} of signal events and \BB{} background events
is obtained from the convolution of Eq.~\ref{eqn:td} with a resolution
function ${\cal R}(\delta t \equiv \deltat -\deltat_{\rm
  true},\sigma_{\deltat})$. The asymmetries \sbb{} and \cbb{} for the
\BB{} background are fixed to zero in the fit, but we account for a
possible deviation from zero in the systematic uncertainty.  The
resolution function is parameterized as the sum of three Gaussian
distributions~\cite{ref:Sin2betaPRD}.  The first two Gaussian
distributions have a width proportional to the reconstructed
$\sigma_{\deltat}$ and a non-zero mean proportional to
$\sigma_{\deltat}$ to account for the small bias in \deltat{} from
charm decays on the \Btag{} side. The third distribution is centered at
zero with a fixed width of \unit[$8$]{ps}.  We have verified in
simulation that the parameters of ${\cal R}(\delta t,
\sigma_{\deltat})$ for \Bztokspizgamma{} events are similar to those
obtained from the \Bflav{} sample, even though the distributions of
$\sigma_{\deltat}$ differ considerably. We therefore extract these
parameters from a fit to the $B_{\rm flav}$ sample. We assume that the
continuum background consists of prompt decays only and find that the
\deltat{} distribution is well described by a resolution function with
the same functional form as used for signal events. The parameters of
the background function are determined in the fit.

\begin{figure}[!tbp]
  \parbox{0.49\linewidth}{
    \includegraphics[width=1.04\linewidth]{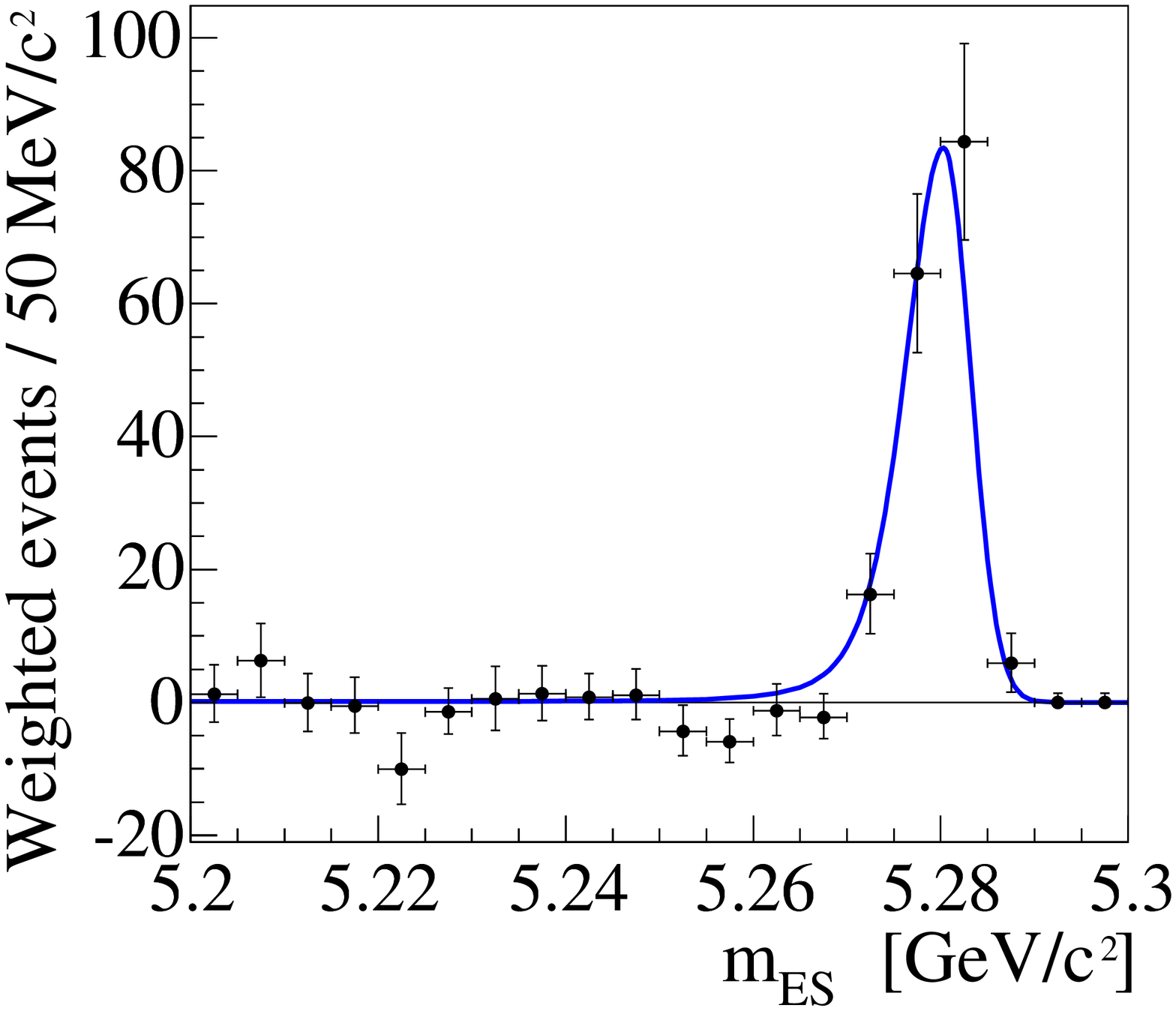}
    }
  \parbox{0.49\linewidth}{
    \includegraphics[width=1.04\linewidth]{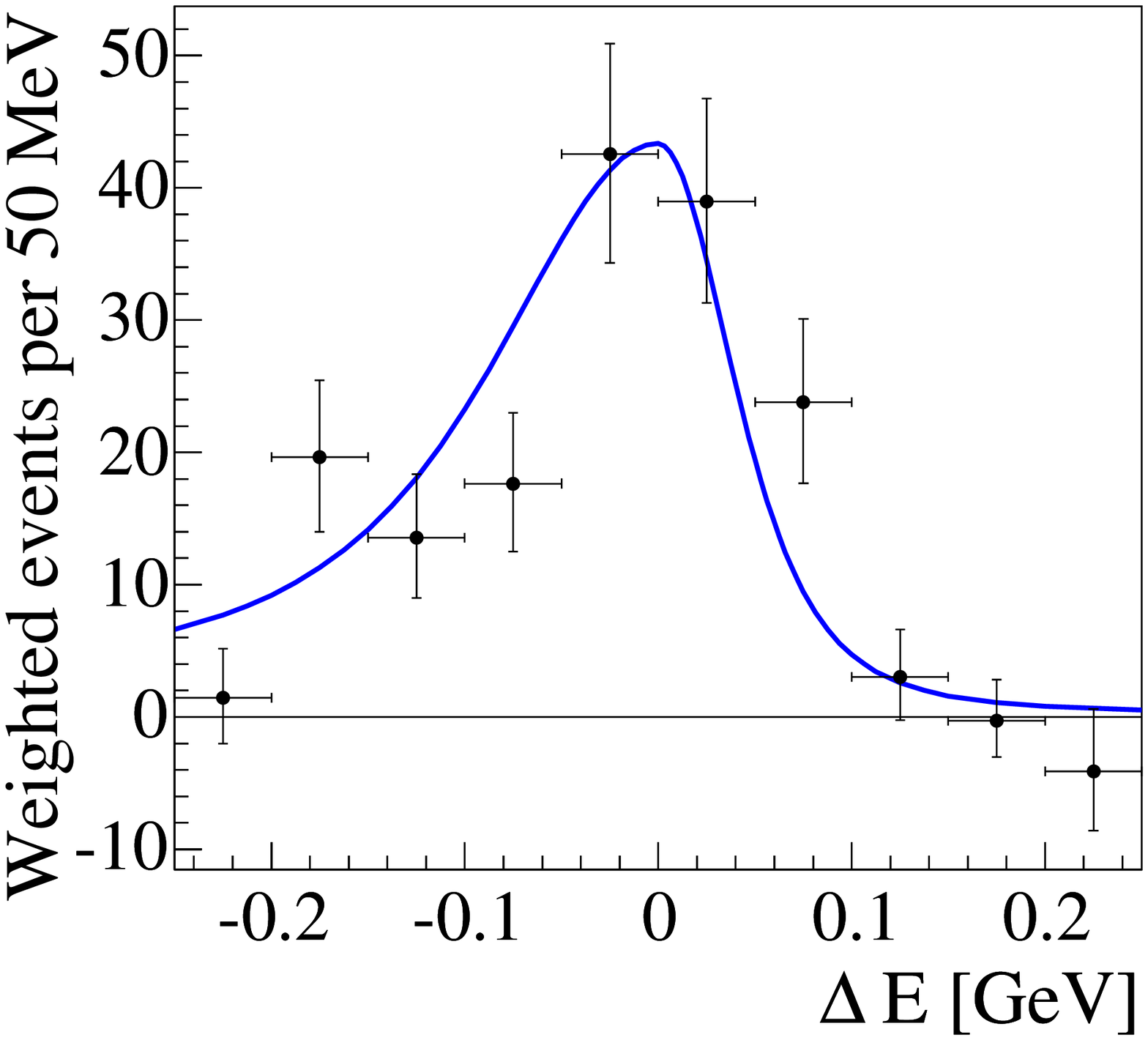}
    }
  \caption{Background-subtracted distribution for \mes{} (left) and \DeltaE{} (right) 
    for $0.8<\mkspiz<1.0$. The lines represent the signal PDFs used in the fit, 
    normalized to the fitted yield.}
  \label{fig:bkinkstar}
\end{figure}

\begin{figure}[!tbp]
  \centerline{\includegraphics[width=0.9\linewidth]{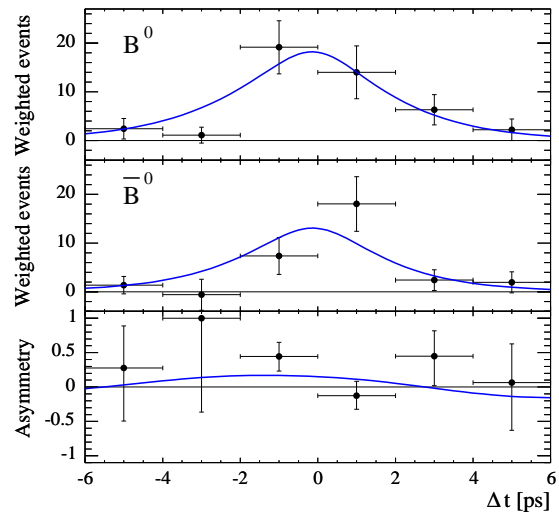}}
  \caption{Background-subtracted distribution for $\deltat$
    with $B_{\rm tag}$ tagged as $\Bz$ (top) or
    $\Bzb$ (center), and the asymmetry ${\cal A}_{\KS\piz}(\deltat)$
    (bottom). The curves represent the PDFs for signal decays in the
    likelihood fit, normalized to the final fit result.}
  \label{fig:dtplot}
\end{figure}

Figure~\ref{fig:bkinkstar} shows the background-subtracted
distributions for \mes{} and \DeltaE{} for the selected
\Bztokstargamma{} candidates. The background subtraction is performed
with the event weighting technique described in~\cite{Pivk:2004ty}.
Events contribute according to a weight constructed from the
covariance matrix for the signal, \BB{} background and continuum
background yields and the probability ${\cal P}_S$, ${\cal P}_{\BB}$
and ${\cal P}_{\qqbar}$ for the event, computed without the use of the
variable that is being displayed.  The curves in the figure represent
the signal PDFs used in the fit. Figure~\ref{fig:dtplot} shows the
background-subtracted distributions of $\deltat$ for $\Bz$- and
$\Bzb$-tagged events, and the asymmetry as a function of $\deltat$.

\begin{figure}[!tbp]
  \centerline{\includegraphics[width=0.7\linewidth]{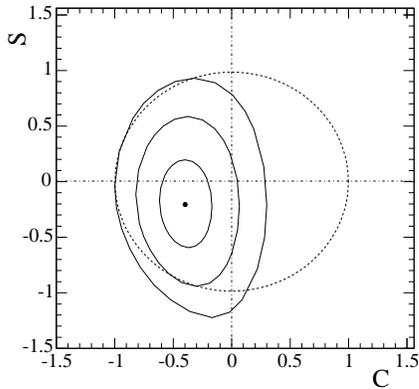}}
  \caption{Constant-likelihood contours in the $S$-$C$ plane for \Bztokstargamma{} corresponding to 
    $-2\Delta\log L = 1$, $2$ and $3$. The dashed circle is the
    physical boundary.}
  \label{fig:contours}
\end{figure}

In the fit to the \Bztokstargamma{} sample we find $157\pm16$ signal
events, with \[\skstargamma=-0.21\pm0.40\pm0.05\] and
\[\ckstargamma=-0.40\pm0.23\pm0.03,\] where the first error is statistical
and the second systematic. The systematic uncertainties are described
below. The extracted \BB{} background is $9\pm13$ events. The linear
correlation coefficient between \skstargamma{} and \ckstargamma{} is
$0.07$. The value of \ckstargamma{} is consistent with the expectation
of no direct \CP{} violation. Since its uncertainty is much larger
than that obtained from the partial rate asymmetry in self-tagging
decays~\cite{HFAGACP}, we also perform the fit with $\ckstargamma$
fixed to zero and find
\[\skstargamma(C\equiv0)=-0.22\pm0.42\pm0.05 .\] The counterintuitive increase in the 
error on \skstargamma{} is a consequence of the likelihood
contours in the $S$-$C$ plane, shown in figure~\ref{fig:contours}, not being
perfectly ellipsoidal.

\begin{figure}[!tbp]
  \centerline{\includegraphics[width=0.7\linewidth]{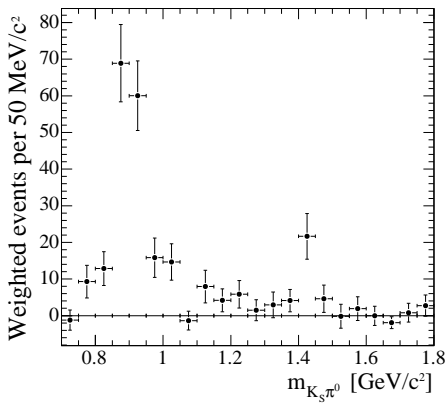}}
  \caption{Background-subtracted distribution for $\mkspiz$. For events with
    \unit[$\mkspiz>1.1$]{\gevcc} the cut on the cosine of the helicity
    angle \cosThetaK{} is not applied.}
  \label{fig:kspizmass}
\end{figure}

\begin{figure}[!tbp]
  \parbox{0.49\linewidth}{
    \includegraphics[width=1.04\linewidth]{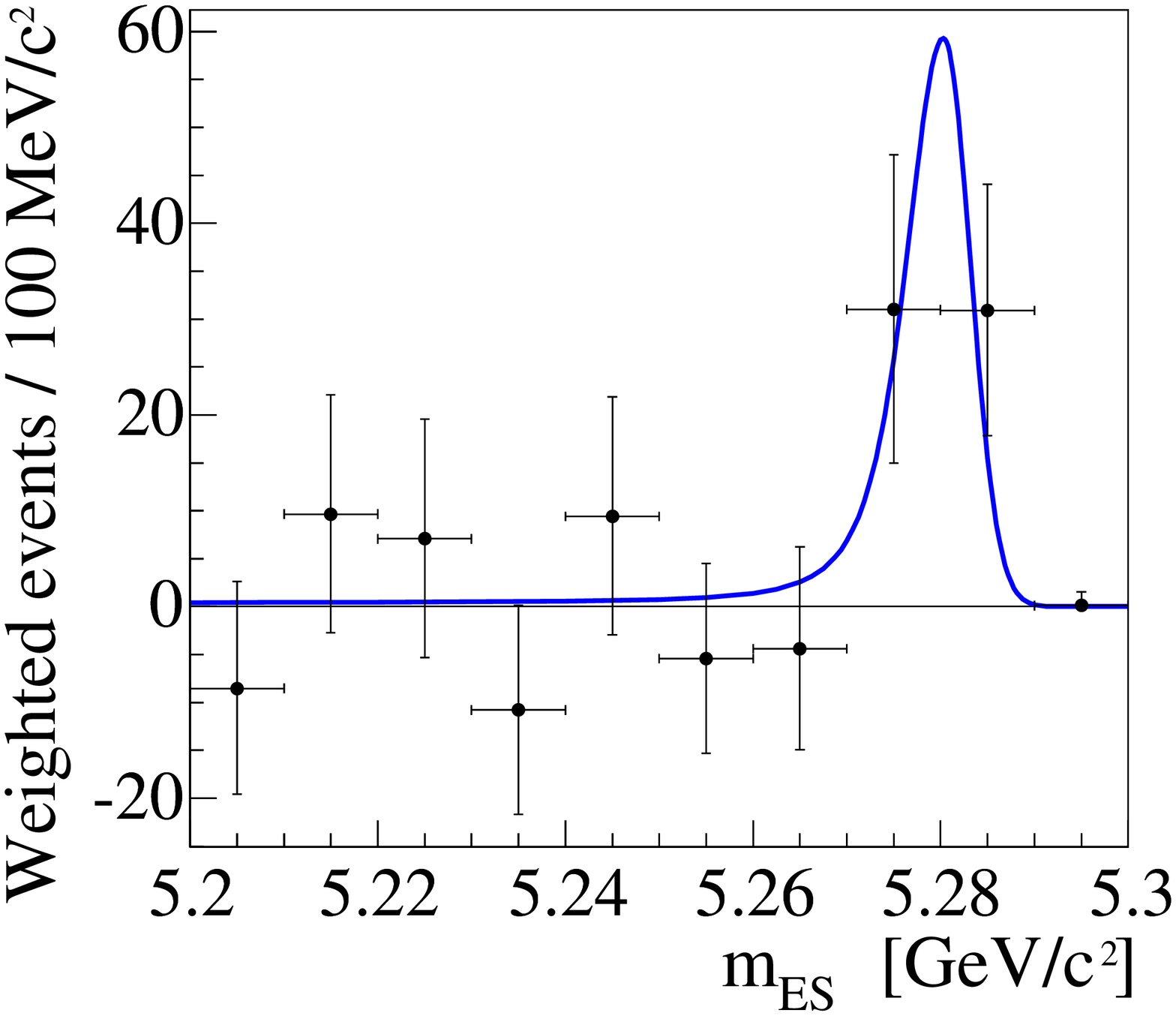}
    }
  \parbox{0.49\linewidth}{
    \includegraphics[width=1.04\linewidth]{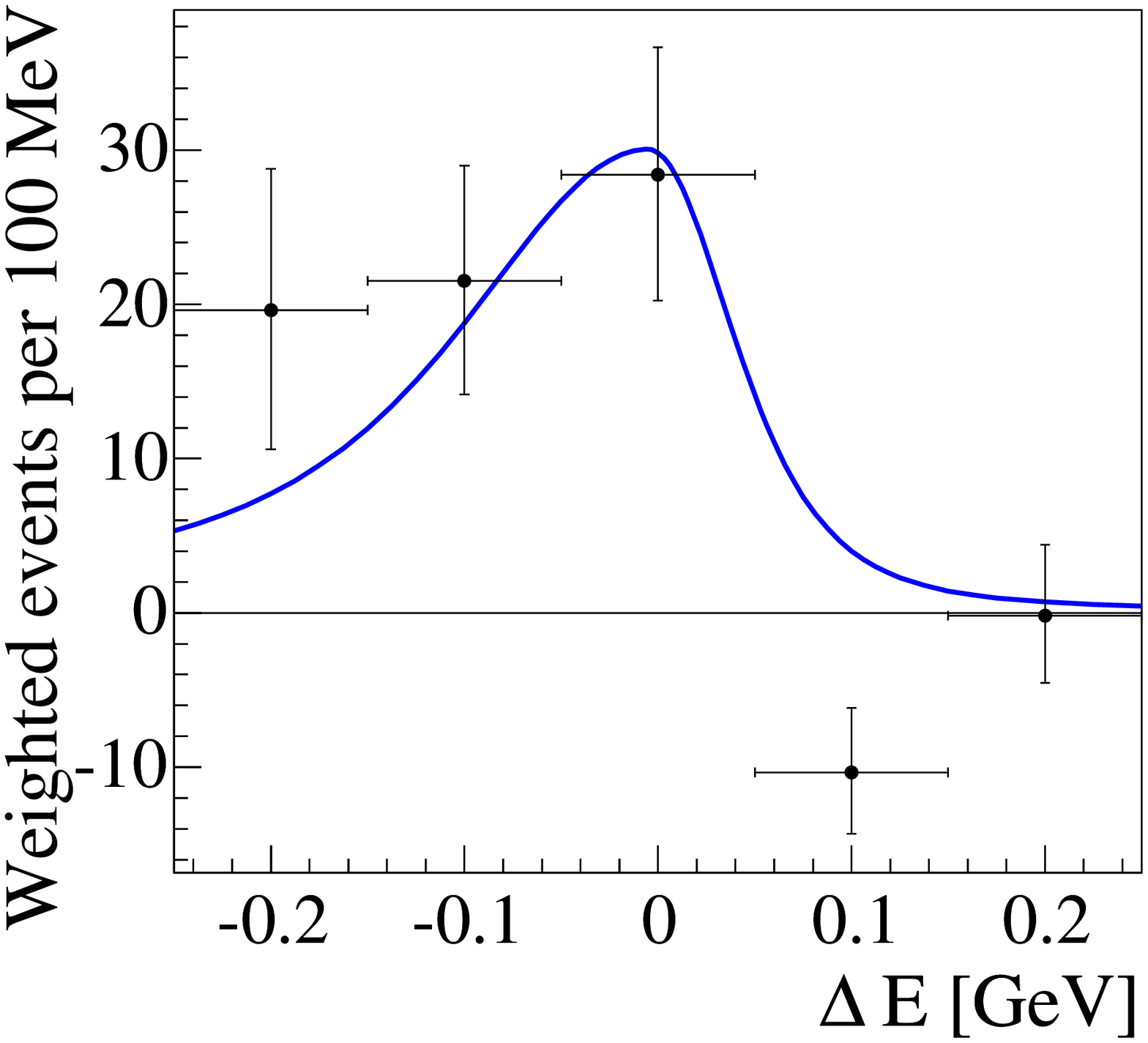}
    }
  \caption{Background-subtracted distribution for \mes{} (left) and \DeltaE{} (right) 
    for $1.1<\mkspiz<1.8$. The lines represent the signal PDFs used in the fit, 
    normalized to the fitted yield.}
  \label{fig:bkinnonres}
\end{figure}

Figure~\ref{fig:kspizmass} shows the background-subtracted $\KS\piz$
invariant mass distribution for $\Bztokspizgamma$ candidates. The
$K^{*}(892)$ resonance is clearly visible and there is some evidence
for the $K^{*}_2(1430)$. Figure~\ref{fig:bkinnonres} shows the
background subtracted distributions for \mes{} and \DeltaE{} events in
the range \unit[$1.1<\mkspiz<1.8$]{\gevcc}.  In the fit to this sample
we find $59\pm13$ signal events with
\[\skspizgamma= 0.9\pm1.0\pm0.2\] and
\[\ckspizgamma=-1.0\pm0.5\pm0.2,\] and $130\pm40$ 
\BB{} background events. The linear correlation coefficient between
\skspizgamma{} and \ckspizgamma{} is $-0.09$.

We consider several sources of systematic uncertainties related to the
level and possible asymmetry of the background contribution from other
\B{} decays.  We evaluate this contribution using simulated samples of
generic $B$ decays and of generic $\B\to\X_s\gamma$ decays. For the
latter we use the Kagan-Neubert model~\cite{Kagan:1998ym} for the
photon energy spectrum and JETSET for the fragmentation of the $s$
quark. Since the final state multiplicity predicted by the
fragmentation model is significantly different from a recent \babar{}
measurement~\cite{ref:babarsumofexclusive}, we reweight events
according to their multiplicity. From these studies we estimate about
$30$ ($140$) events in the $K^*$ (non-$K^*$)  sample, with approximately equal
contributions $\B\to\X_s\gamma$ decays and other (generic) $B$ decays.
In the $K^*$ sample we fit a
contribution from \BB{} background consistent both with zero and with
the rate predicted from the simulation. In the non-$K^*$ sample 
 the fitted contribution from other
\B{} decays is significantly larger, as expected from the simulation.
Using the Monte Carlo estimates for the yields, we assess the impact
of a potential CPV asymmetry in the $\BB$ background by varying \sbb{}
and \cbb{} within an appropriate range that is derived from the
composition of the $\BB$ background sample. We assign a systematic
uncertainty of $0.04$ ($0.03$) on $S$ ($C$) in the $K^*$ 
sample and an uncertainty of $0.2$
for both $S$ and $C$ in the non-$K^*$ sample.

We quantify possible systematic effects due to the vertex
reconstruction method in the same manner as in
Ref.~\cite{Aubert:2004xf}, estimating systematic uncertainties on $S$
($C$) of $0.023$ ($0.014$) due to the vertex reconstruction technique
and uncertainties in the resolution function, and $0.020$ ($0.007$)
due to possible misalignments of the SVT.  Finally, we include a
systematic uncertainty due to imperfect knowledge of the PDFs used in
the fit, which amounts to $0.02$ ($0.01$) for the $K^*$ (non-$K^*$)
sample.

In summary, we have performed a new measurement of the time-dependent
CPV asymmetry in \Bztokstargamma{} decays. Within the large
statistical uncertainties our measurement is consistent with the SM
expectation of a small CPV asymmetry and with other
measurements~\cite{Ushiroda:2005sb}. We have also explored the
possibility of measuring the CPV asymmetry in the region with a
$\KS\piz$ invariant mass above the \Kstarz{} region,
\unit[$1.1<\mkspiz<1.8$]{\gevcc}. We find that the signal yield,
though consistent with the expectation, is too small for a meaningful
asymmetry measurement. These results supersede our previous
measurement~\cite{Aubert:2004pe} which was based on a subset of the
data presented here.

We are grateful for the 
extraordinary contributions of our \pep2\ colleagues in
achieving the excellent luminosity and machine conditions
that have made this work possible.
The success of this project also relies critically on the 
expertise and dedication of the computing organizations that 
support \babar.
The collaborating institutions wish to thank 
SLAC for its support and the kind hospitality extended to them. 
This work is supported by the
US Department of Energy
and National Science Foundation, the
Natural Sciences and Engineering Research Council (Canada),
Institute of High Energy Physics (China), the
Commissariat \`a l'Energie Atomique and
Institut National de Physique Nucl\'eaire et de Physique des Particules
(France), the
Bundesministerium f\"ur Bildung und Forschung and
Deutsche Forschungsgemeinschaft
(Germany), the
Istituto Nazionale di Fisica Nucleare (Italy),
the Foundation for Fundamental Research on Matter (The Netherlands),
the Research Council of Norway, the
Ministry of Science and Technology of the Russian Federation, and the
Particle Physics and Astronomy Research Council (United Kingdom). 
Individuals have received support from 
CONACyT (Mexico),
the A. P. Sloan Foundation, 
the Research Corporation,
and the Alexander von Humboldt Foundation.


\begin{thebibliography}{99}
\bibitem{ref:cc}
Unless explicitly stated, charge conjugate decay modes are included 
implicitly throughout this paper.
\end{thebibliography}
\end{document}